\documentclass[twocolumn,showpacs,showkeys,amsmath,amssymb]{revtex4}

\usepackage{graphicx}

\begin{document}

\title{A comparison of approximate gravitational lens equations and
 a proposal for an improved new one}

\author{V. Bozza$^{a,b}$}

\affiliation{$^a$ Dipartimento di Fisica ``E.R. Caianiello'',
Universit\`a di Salerno, via Allende,
I-84081 Baronissi (SA), Italy.\\
$^b$ Istituto Nazionale di Fisica Nucleare, Sezione di Napoli.}

\date{\today}

\begin{abstract}
Keeping the exact general relativistic treatment of light bending
as a reference, we compare the accuracy of commonly used
approximate lens equations. We conclude that the best approximate
lens equation is the Ohanian lens equation, for which we present a
new expression in terms of distances between observer, lens and
source planes. We also examine a realistic gravitational lensing
case, showing that the precision of the Ohanian lens equation
might be required for a reliable treatment of gravitational
lensing and a correct extraction of the full information about
gravitational physics.
\end{abstract}



\maketitle

\section{Introduction}

Gravitational lensing is a well-established research subject
treating the bending of light trajectories by gravitational
fields. Its methodology is traditionally developed within the weak
field approximation of General Relativity, which describes photon
trajectories through the geodesics equation. In this context, the
true position of a source in the sky and its apparent position
after deflection of the light by a massive body are related by the
so-called lens equation \cite{Liebes,Refsdal} (for textbook
reviews, see \cite{SEF,PLW,MolRou}). This relation and its
mathematical properties have been extensively studied for generic
lens models and represents the basis of the whole gravitational
lensing theory.

The lens equation is typically established in the small angle
approximation, in conjunction with the weak deflection hypothesis.
In more recent years, there has been a renewed interest in
astrophysical situations in which the deflection angle is not
small. For example, if the bending of light emitted by sources
near a black hole is considered, the deflection angle may reach
arbitrary large values \cite{Dar}. For such gravitational lenses
the old small angles lens equation must be obviously revised.
Several proposals for generalized lens equations have then
appeared in the literature and have been applied to specific cases
\cite{FriNew,FKN,Oha,VNC,VirEll,DabSch,BozSer}. Such lens
equations principally differ among each other for what concerns
the variables in which they are expressed. However they also lie
at different approximation levels below the full general
relativistic description of the photon motion.

In this work, we review all these lens equations for spherically
symmetric bodies appeared in the literature and introduce a new
lens equation that fills a gap in the present taxonomy. We also
present a detailed discussion of the order of magnitude of the
errors committed in the use of different lens equations, giving a
complete interpretation for their origin. It is worth mentioning
that a critical review of the approximations leading to the weak
deflection lens equation has recently appeared \cite{KliFri}. The
present work shares the same spirit by testing different
approximations leading to large deflection lens equations,
instead.

The plan of the paper is as follows. In Section II we establish
the notations by describing the basic geometric configuration for
gravitational lensing. In Section III, we review the exact lens
equation by Frittelli, Kling and Newman, derived in a fully
general relativistic context. In Section IV, we discuss the
asymptotic approximation and all lens equations making use of this
approximation. We also introduce our new proposal for an improved
lens equation. In Section V we present a numerical example about a
realistic gravitational lensing situation, in order to compare the
different approximations in the lens equations previously
discussed. Section VI contains a discussion about the precision
needed in gravitational lensing observations and the conclusions
of the work.

\section{Basic lensing geometry}

Let us consider a spherically symmetric spacetime, whose metric is
\begin{equation}
ds^2=A(r)dt^2-B(r)dr^2-C(r)^2(d\vartheta^2+\sin \vartheta^2
d\phi^2). \label{Metric}
\end{equation}

Let us put a static source at radial coordinate $d_{LS}$ and a
static observer at radial coordinate $d_{OL}$. The orientation of
the polar coordinates is chosen in such a way that both the source
and the observer lie on the equatorial plane $\vartheta=\pi/2$. As
a consequence of the spherical symmetry, the whole photon motion
takes place on this plane. We also assume that the metric is
asymptotically flat, so that the coordinates
$(t,r,\vartheta,\phi)$ become Minkowski polar coordinates very far
from the center (we neglect any effects due to cosmological
expansion in our discussion).

\begin{figure*}
\resizebox{\hsize}{!}{\includegraphics{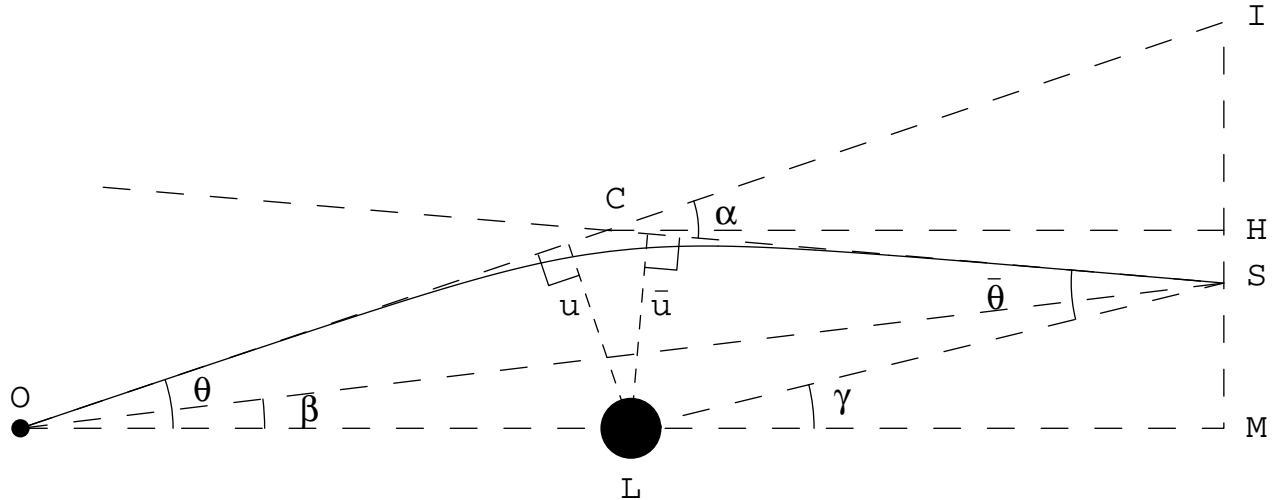}} \caption{Generic
gravitational lensing configuration. Note that only the angles
$\gamma$, $\theta$ and $\bar\theta$ are well-defined in the curved
spacetime, whereas all other geometrical quantities need to be
defined referring to the Minkowski space associated with the
original curved space.}
 \label{Fig}
\end{figure*}

A very useful structure for understanding several lens equations
proposed in the literature is the Minkowski space associated with
(\ref{Metric}). This structure can be introduced by assuming that
the metric depends on one or more parameters continuously so that,
once we tune these parameters to zero, the metric becomes
Minkowski and the coordinates $(t,r,\vartheta,\phi)$ become
Minkowski polar coordinates on the whole spacetime. The parameters
of the metric may be the mass of the lens, the electric charge,
the scalar charge or any other parameter dictated by any kind of
gravitational theory. Many definitions need to be given by
referring to this associated Minkowski space obtained by tuning
all these parameters to zero. Of course, the original curved
spacetime and the associated Minkowski space coincide in the
asymptotically flat region.

A typical geometric configuration of gravitational lensing is
shown in Fig. \ref{Fig}, with a photon emitted by a source $S$,
curved by the gravitational field generated by the lens $L$ and
detected by the observer in $O$. The observer sees the photon at
an angle $\theta$ with respect to the optical axis $OLM$, whereas,
if there were no lens, he would directly observe the source at an
angle $\beta$. $\beta$ is thus the first quantity that is
well-defined only in the associated Minkowski space, as we must
tune the mass of the lens to zero to define it.

The emission direction of the photon is $SC$, whereas the
detection direction is $CO$. In the associated Minkowski space, we
can also measure the angle between these two directions and define
it as the deflection angle $\alpha$ (we cannot compare directions
from different points in a curved space).

It is common practice to define the lens plane and the source
plane as those planes orthogonal to the optical axis passing
through the lens and the source, respectively. Of course, also
these definitions can be given in the associated Minkowski space
and then extended to the original curved space. Then, one can
define the distances from the observer to the lens plane as
$D_{OL}$, the distance between the lens plane and the source plane
as $D_{LS}$ and the distance between the observer and the source
plane as $D_{OS}$. The usual relation
\begin{equation}
D_{OS}=D_{OL}+D_{LS}
\end{equation}
holds.

These distances between planes are obviously different from the
distances between the points $O$, $L$ and $S$. We have already
defined $d_{OL}$ and $d_{LS}$ as the radial coordinates of
observer and source in the original curved metric. Once we report
these coordinates in the associated Minkowski space, they coincide
with the proper distances of the observer and the source from the
lens. Then, in this space, we can establish simple geometrical
relations with the distances between planes. In particular, we
have
\begin{eqnarray}
&& d_{OL}=D_{OL} \\
 && d_{LS}=D_{LS}/\cos{\gamma} \label{dD1}
\\ && d_{OS}=D_{OS}/\cos{\beta}. \label{dD2}
\end{eqnarray}

Of course, if the source is very close to the optical axis,
$\beta$ and $\gamma$ are small and the differences between the
uppercase distances and the lowercase distances is of second order
in the angles. Therefore, in the classical weak deflection
paradigm, the two notions are confused without consequences. Here,
we must keep them distinct in order to avoid confusion.

\section{Exact lens equation}

The lens equation is a relation among the source and observer
coordinates and the angle $\theta$ at which the observer detects
an image of the source $S$.

Remaining in a fully general relativistic context, it is possible
to write down the exact equations governing the photon motion and
consequently write an exact lens equation. This approach has been
proposed by Frittelli and Newman in Ref. \cite{FriNew} and then
applied to the Schwarzschild lens in Ref. \cite{FKN}. Later on, it
has been generalized to all spherically symmetric spacetimes in
Ref. \cite{Perlick} (see also \cite{LivRev}).

A photon emitted from a source, deflected by the lens and detected
by the observer experiences a change in the azimuthal coordinate
given by
\begin{equation}
\Phi(J,d_{OL},d_{LS})=\left[\int\limits_{r_0}^{d_{OL}}+\int\limits_{r_0}^{d_{LS}}\right]
\frac{\sqrt{B(r)}J}{C(r)\sqrt{\frac{1}{A(r)}-\frac{J^2}{C(r)}}}dr,
\label{Phi}
\end{equation}
where $J$ is the specific angular momentum of the photon, which is
a constant of motion, and $r_0$ is the distance of closest
approach to the lens. These two quantities are related by
\begin{equation}
J=\sqrt{\frac{C(r_0)}{A(r_0)}}. \label{Jr0}
\end{equation}

The specific angular momentum is also related to the angle
$\theta$ at which the observer detects the photon. Defining the
angle by using the scalar product between the arrival direction
and the direction specified by the optical axis $OL$ (see e.g.
\cite{Nem}), we find
\begin{equation}
\theta=\arcsin \left(J\sqrt{\frac{A(d_{OL})}{C(d_{OL})}} \right).
\label{thetaJ}
\end{equation}

If the observer is very far from the lens, $A(d_{OL}) \rightarrow
1$ and $C(d_{OL}) \rightarrow d_{OL}^2$, so that one recovers the
relation
\begin{equation}
\theta \simeq \arcsin J/d_{OL}, \label{thetaJapp}
\end{equation}
which allows to identify $J$ with the impact parameter $u$ of the
light ray trajectory (except for a speed of light factor). In
general, however, this identification is only approximate (see
next section).

The exact lens equation can finally be written by noting that the
change in the azimuthal coordinate must be equal to the difference
between the azimuthal coordinates of the observer and the source.
Replacing all occurrences of $J$ and $r_0$ in terms of $\theta$ by
Eqs. (\ref{Jr0}) and (\ref{thetaJ}), we get
\cite{FriNew,FKN,Perlick}
\begin{equation}
\Phi(\theta,d_{OL},d_{LS})=\pi-\gamma. \label{LensFKN}
\end{equation}

This equation is the exact general relativistic relation between
the angle at which the image appears in the observer's sky and the
relative positions of source, observer and lens. All other lens
equations represent approximate forms of this equation. Of course,
most lens equations are derived under assumptions that are largely
satisfied in realistic situations. Therefore, their
simplifications are absolutely welcome, if the induced errors are
below observational sensitivity or other sources of noise. In the
following sections we will introduce some lens equations,
describing their approximations.

\section{Approximate lens equations}

The most popular approximation is what can be called as asymptotic
approximation, which amounts to saying that the source and the
observer are in the asymptotic flat region of the spacetime.
Quantitatively, this is expressed by requiring $d_{OL},d_{LS} \gg
r_g$, where $r_g$ is the gravitational radius of the lens, i.e.
the typical scale in the curved spacetime metric controlling the
range of the gravitational field (it is $2GM/c^2$ in the
Schwarzschild metric). As a first consequence, Eq.
(\ref{thetaJapp}) holds, so that the angular momentum of the
photon $J$ can be identified with the impact parameter $\bar u$ of
the initial trajectory and the impact parameter $u$ of the final
trajectory, as depicted in Fig. \ref{Fig}. As a second
consequence, the deflection angle can be calculated as
\begin{equation}
\alpha(u)\equiv\Phi(J=u,\infty,\infty)-\pi, \label{alpha}
\end{equation}
which is the azimuthal shift of a photon incoming from infinity
with impact parameter $J$ and escaping to infinity.

Once this approximation is accepted, it is possible to establish a
relation among $\theta$ and the relative positions of source, lens
and observer using pure Euclidean geometry. In fact, all distances
are defined in the asymptotic spacetime and the only input from
General Relativity is the precise expression of the deflection
angle as a function of the impact parameter $u$.

Note that the asymptotic approximation does not necessary imply
the small angle approximation. In fact, it just assumes that the
source and the observer distances are much larger than the
gravitational radius of the lens. This does not prevent the impact
parameter to be very large.

In the following subsections, we will introduce several
approximate lens equations appeared in the literature, expressed
in terms of different quantities.

\subsection{Ohanian lens equation}

The first lens equation we introduce is due to Ohanian \cite{Oha},
who proposed it in a study of gravitational lensing by a
Schwarzschild black hole for arbitrary deflection angles.

Thanks to the asymptotic approximation, we can use Euclidean
geometry to relate the various quantities and store all the
relativistic input in the angle $\alpha$. Let us define the angle
$\zeta\equiv \widehat{OSL}$. Considering the triangles $OLS$ and
$OSC$, we can write down the relations
\begin{eqnarray}
&& \beta+\zeta+\pi-\gamma=\pi \\
&& (\theta-\beta)+(\bar\theta-\zeta)+\pi-\alpha=\pi.
\end{eqnarray}
Summing up these two equalities, we obtain \cite{BozMan} (see also
\cite{BozSer})
\begin{equation}
\theta+\bar\theta-\alpha=\gamma. \label{LensOha}
\end{equation}

The angle $\bar\theta$ can be expressed in terms of $\theta$
recalling that the impact parameter of the incoming trajectory
$\bar u$ is equal to the impact parameter of the outgoing
trajectory $u$ (in the asymptotic approximation). Therefore
\begin{equation}
\bar\theta=\arcsin\left(\frac{d_{OL}}{d_{LS}} \sin\theta \right).
\label{bartheta}
\end{equation}
Once Eq. (\ref{bartheta}) is used in Eq. \ref{LensOha}, we obtain
the lens equation as a relation involving the detection angle
$\theta$, the distance between the observer and the lens $d_{OL}$,
the distance of the lens to the source $d_{LS}$, the source
position angle $\gamma$ and the deflection angle $\alpha$, which
is a function of $\theta$ through $u=d_{OL}\sin \theta$.

The original form proposed by Ohanian \cite{Oha} was actually
written by replacing $\theta$ and $\bar\theta$ by their small
angle approximations $u/d_{OL}$ and $u/d_{LS}$ respectively. This
is an additional approximation, with respect to the asymptotic
approximation, which we will not consider here.

\subsection{VNC lens equation}

Another very simple expression for the lens equation was proposed
by Virbhadra, Narasimha and Chitre (VNC) in a paper studying the
role of the scalar field in gravitational lensing \cite{VNC}. It
is just the result of the sine theorem applied to the triangle
$OCS$
\begin{equation}
\sin(\theta-\beta)=\frac{d_{CS}}{d_{OS}}\sin \alpha,
\label{LensVNC}
\end{equation}
where $d_{CS}$ is the distance between the source $S$ and the
intersection point $C$ between the incoming and outgoing
trajectories. Eq. (\ref{LensVNC}) is valid for arbitrary
deflections and impact parameters. As for the Ohanian lens
equation, its only approximation with respect to the exact lens
equation (\ref{LensFKN}) is the asymptotic approximation, which is
necessary to express the deflection angle $\alpha$.

However, the lens equation (\ref{LensVNC}) is not a closed
relation among the detection angle $\theta$ and the relative
positions of observer, lens and source. In fact, it is expressed
in terms of the position of point $C$, which is unknown a priori
and needs to be estimated in some way. One possibility is to
approximate $d_{CS}$ by $d_{LS}$, which is reasonable if the
source is at distances much larger than the impact parameter.
However, this additional approximation would spoil the
effectiveness of the lens equation in its original form.
Therefore, the VNC lens equation is difficult to use in practical
applications, since it demands an independent knowledge or at
least an estimate of the position of the point $C$.

\subsection{Virbhadra and Ellis lens equation} \label{Sec VE}

With their outbreaking paper about the possibility of observing
higher order images around the black hole at the center of our
Galaxy \cite{VirEll}, Virbhadra and Ellis have attracted a great
attention on gravitational lensing beyond the weak deflection
approximation, inspiring new vitality in black hole gravitational
lensing. They have also proposed a new lens equation that has
become very popular in the scientific literature. Their equation
is written in terms of the distances between source, lens and
observer planes.

Starting from the relation among the segments
\begin{equation}
\overline{MS}=\overline{MI}-\overline{SI},
\end{equation}
we can write the relation
\begin{equation}
D_{OS}\tan \beta = D_{OS} \tan{\theta} - D_{CS}
\left[\tan\theta+\tan (\alpha - \theta) \right],
\end{equation}
where $D_{CS}$ is the distance between point $C$ and the source
plane (namely the length of the segment $\overline{CH}$). Even in
this case, the lens equation is expressed in terms of the position
of point $C$, which should be estimated in some way. The proposal
by Virbhadra and Ellis is to assume that $C$ lies on the lens
plane, so that $D_{CS}\simeq D_{LS}$. Then the final form of the
lens equation is
\begin{equation}
D_{OS}\tan \beta = D_{OS} \tan{\theta} - D_{LS}
\left[\tan\theta+\tan (\alpha - \theta) \right], \label{LensVE}
\end{equation}
At the cost of an additional approximation, the lens equation is
put in a form very easy to use in many astrophysical applications,
as it is finally expressed in terms of the positions of lens,
source and observer. We will discuss the error introduced by this
approximation in Section \ref{Sec Num}, while in Section \ref{Sec
Boz} we will present an improvement of this equation which avoids
this approximation and makes it equivalent to the Ohanian lens
equation.

\subsection{Dabrowski and Schunck lens equation}

In a paper studying gravitational lensing by boson stars
\cite{DabSch}, Dabrwoski and Schunck realized the difficulties of
using the VNC lens equation (\ref{LensVNC}) and derived the
alternative lens equation

\begin{eqnarray}
&\sin (\theta -\beta)&=\frac{d_{LS}}{d_{OS}} \cos \theta \cos
\left\{ \arcsin \left[\frac{d_{OS}}{d_{LS}} \sin\beta \right]
\right\} \nonumber \\ && \times
\left[\tan\theta+\tan(\alpha-\theta) \right]. \label{LensDS}
\end{eqnarray}

This can be obtained from the Virbhadra and Ellis lens equation
(\ref{LensVE}) replacing the distances between planes $D_{OS}$ and
$D_{LS}$ by the distances between objects using Eqs. (\ref{dD1})
and (\ref{dD2}) and noting that
\begin{equation}
d_{LS} \sin \gamma = d_{OS} \sin\beta \label{gammabeta}
\end{equation}
in the associated Minkowski space.

In practice, the equation (\ref{LensDS}) by Dabrwoski and Schunck
is the same as Eq. (\ref{LensVE}) by Virbhadra and Ellis but
expressed in terms of distances between objects instead of
distances between planes. Of course, like the former equation, it
contains the additional approximation that $C$ lies on the lens
plane.

\subsection{Bozza and Sereno lens equation}

The Ohanian lens equation (\ref{LensOha}) has the advantage of
being very simple and being the closest relative of the exact lens
equation, since it only contains the asymptotic approximation and
makes no additional assumptions. However, in several astrophysical
applications one may prefer to have a lens equation directly
written in terms of the angle $\beta$ rather than $\gamma$. In
fact, $\beta$ is an angle with vertex in the observer and thus
directly connected to coordinates in the observer's sky, whereas
$\gamma$ is an angle with vertex in the lens, thus being less
close to observables. The relation between the two angles is given
by Eq. (\ref{gammabeta}). Therefore, if we take the sine of
equation (\ref{LensOha}), using Eq. (\ref{bartheta}) and
re-ordering terms we get
\begin{eqnarray}
&d_{OS} \sin \beta =& d_{OL} \sin \theta \cos(\alpha
-\theta)\nonumber \\&&-\sqrt{d_{LS}^2-d_{OL}^2\sin^2\theta}
\sin(\alpha -\theta), \label{LensBS}
\end{eqnarray}
which first appeared in Ref. \cite{BozSer}. Though being more
complicated than Eq. (\ref{LensOha}), this lens equation has the
advantage of being directly expressed in terms of the angle
$\beta$ and thus preferable for applications in which this angle
is directly involved. Otherwise, it is completely equivalent to
the Ohanian lens equation. Note that the distances involved in
this equation are distances between the objects and not between
their geometrical planes, as in the lens equations of Sections
\ref{Sec VE} and \ref{Sec Boz}.

One may note that this lens equation is expressed in terms of the
same variables appearing in the equation by Dabrwoski and Schunck
(\ref{LensDS}). However the two equations look different. Apart
from the ordering of the terms, Eq. (\ref{LensDS}) contains the
additional approximation that $C$ lies on the lens plane, whereas
Eq. (\ref{LensBS}) does not. Therefore, Eq. (\ref{LensBS})
represents an improved version of Eq. (\ref{LensDS}). The same
difference will arise in the new equation to be presented in the
next section and the equation by Virbhadra and Ellis.

\subsection{A new improved lens equations between planes} \label{Sec
Boz}

Eq. (\ref{LensVE}) represents a very useful lens equation
expressed in terms of distances between the observer, lens and
source planes. However, as we have explained in Section \ref{Sec
VE}, it is derived under the additional assumption that the
intersection point $C$ between the incoming and outgoing ray
trajectories lies on the lens plane. It would be desirable to have
a lens equation expressed in terms of the same quantities without
this additional assumption.

Starting again from the Ohanian lens equation (\ref{LensOha}), we
can solve it in terms of $\bar\theta$ and plug it in Eq.
(\ref{bartheta}). Recalling Eq. (\ref{dD1}) and the relation
between $\gamma$ and $\beta$ (\ref{gammabeta}), we can put the
lens equation in the form
\begin{equation}
D_{OS} \tan \beta=\frac{D_{OL}\sin \theta - D_{LS}
\sin(\alpha-\theta)}{\cos(\alpha -\theta)}, \label{LensEq}
\end{equation}
which represents the improved version of Eq. (\ref{LensVE}) by
Virbhadra and Ellis, in the same sense as Eq. (\ref{LensBS}) by
Bozza and Sereno represents the improved version of Eq.
(\ref{LensDS}) by Dabrowski and Schunck. In Section V we make a
thorough discussion and comparison of the two equations.

\subsection{Small angles lens equation} \label{Sec SA}

We finally recall the classical lens equation obtained in the
hypothesis of small angles $\alpha,\theta,\beta \ll 1$. This
hypothesis has nothing to do with the weak deflection
approximation, since the deflection angle is typically expressed
in powers of $r_g/u$, whereas the corrections to the small angles
approximation are expressed in powers of $\theta,\alpha,\beta$. So
it makes sense to consider an exact deflection angle while
performing the small angles approximation in all trigonometric
functions.

The small angles lens equation can be obtained from any of the
equations (\ref{LensVE}), (\ref{LensDS}), (\ref{LensBS}) or
(\ref{LensEq}). The well-known result
\cite{Liebes,Refsdal,SEF,PLW,MolRou} is
\begin{equation}
\beta=\theta - \frac{D_{LS}}{D_{OS}} \alpha. \label{LensSA}
\end{equation}

The small angles approximation is the most rude approximation that
can be done on the lens equation. Nevertheless, it is a useful
reference approximate equation lying on the other extremum of the
approximation ladder with respect to the exact lens equation
(\ref{LensFKN}), with all other lens equations staying in the
middle steps between the two equations.

It is interesting to note that a similar equation can be deduced
when the deflection angle is very close to multiples of $2\pi$.
This case corresponds to higher order images generated by photons
performing one or more loops around the lens before emerging. The
only difference is that the deflection angle $\alpha$ in Eq.
(\ref{LensSA}) must be replaced by $\alpha-2n\pi$, where $n$ is
the number of loops.

\subsection{Conclusions on approximate lens equations}

To summarize this section, we can say that the asymptotic
approximation allows the use of the relations of Euclidean
geometry, confining General Relativity to the derivation of the
precise expression of the deflection angle in terms of the impact
parameter.

In order to write the lens equation in terms of $\theta$, $\alpha$
and the relative positions of observer, source and lens, an
additional relation is needed between the incoming and the
outgoing branches of the photon trajectory. The lens equation by
Ohanian (\ref{LensOha}), and the related equations (\ref{LensBS})
by Bozza and Sereno and (\ref{LensEq}) presented in this paper,
use the equality of the impact parameters of the incoming and
outgoing trajectories, expressed by Eq. (\ref{bartheta}). This
equality descends from the time-reversal symmetry of the photon
geodesic in General Relativity along with the asymptotic
approximation.

On the other hand, the lens equations by Virbhadra and Ellis
(\ref{LensVE}) and Dabrowski and Schunck (\ref{LensDS}) use an
alternative relation between the incoming and outgoing
trajectories by imposing that the intersection point $C$ lies on
the lens plane. In the next section we will quantify the accuracy
of these lens equations in a realistic astrophysical situation.

\section{Lens equations at work in a realistic example} \label{Sec Num}

As explained in the previous section, in the treatment of
gravitational lensing by spherically symmetric lenses with
arbitrary deflection angles, we can distinguish an exact lens
equation (\ref{LensFKN}), a family of lens equations with the
asymptotic approximation only (\ref{LensOha}), (\ref{LensBS}) and
(\ref{LensEq}), and a family of lens equations with the additional
geometrical approximation that the intersection point $C$ lies on
the lens plane (\ref{LensVE}) and (\ref{LensDS}). For simplicity,
we shall refer to these two families of approximate lens equations
as the Ohanian family and the Virbhadra and Ellis family,
respectively. Furthermore, we have the small angles lens equation,
in which all trigonometric functions are approximated by their
first order expansions. Finally, we can also consider the effect
of the weak deflection approximation on the deflection angle, by
retaining only the lowest order term in the weak deflection
expansion. We will not consider here the VNC lens equation
(\ref{LensVNC}) since it is not expressed in terms of the relative
configuration of source, lens and observer but demands additional
information on the position of point $C$.

As pointed out in Ref. \cite{VirEll}, the best candidate for the
observation of gravitational lensing in the regime of large
deflections is the black hole at the center of our Galaxy, named
Sgr A*. Its mass is estimated to $M=3.6\times 10^6$ $M_\odot$ and
its distance is $D_{OL}=8$ kpc \cite{Eis}. The Schwarzschild
radius is therefore
\begin{equation}
r_{g}=\frac{2GM}{c^2}=10^{10} m. \label{rg}
\end{equation}

In order to test the accuracy of the different lens equations, we
consider the source plane to be at distance $D_{LS}=1000 r_g$ from
the black hole. At this distance, the source is just at the margin
of the accretion disk of the black hole, so it may be a star
spiralling towards the black hole (the star S2 is known to reach a
distance of 1700$r_g$ at its periapse \cite{S2,Eis}) or a bright
spot of hot material rotating around the black hole. Apart from
the physical meaning of a source at such distance, this choice has
been made with the aim of putting in better numerical evidence the
differences in the performances of the lens equations, since the
expected order of magnitude of the errors scales with powers of
$r_g/D_{LS}$ and $r_g/D_{OL}$.

We are going to study the position of the images as a function of
the position of the source, keeping the distance to the source
plane $D_{OS}$ fixed. Therefore, we will compare the results of
the exact lens equation (\ref{LensFKN}) with the results of the
Virbhadra and Ellis (VE) lens equation (\ref{LensVE}), the
improved lens equation (\ref{LensEq}), which is equivalent to
Ohanian lens equation (we shall refer to it as the OB lens
equation), and the small angles lens equation (\ref{LensSA}). The
last three equations are already expressed in terms of $\beta$,
$D_{LS}$ and $D_{OS}$, whereas the exact lens equation
(\ref{LensFKN}) is expressed in terms of $\gamma$ and $d_{LS}$. We
will use Eqs. (\ref{gammabeta}) and (\ref{dD1}) to obtain these
quantities as functions of $\beta$ and $D_{LS}$, referring to the
associated Minkowski spacetime.

The azimuthal shift $\Phi$ is calculated assuming that the
spacetime metric around the black hole is Schwarzschild. Therefore
we have
\begin{equation}
A(r)=\left(1-\frac{r_g}{r} \right); \;\;
B(r)=\left(1-\frac{r_g}{r} \right)^{-1}; \;\; C(r)=r^2.
\end{equation}
We remind that the detection angle $\theta$ is related to the
angular momentum of the photon $J$ by Eq. (\ref{thetaJ}). In the
approximate lens equations (\ref{LensVE}), (\ref{LensEq}) and
(\ref{LensSA}) the deflection angle is obtained by Eq.
(\ref{alpha}) as a function of $\theta$ through Eq.
(\ref{thetaJ}). For any given value of the source position
$\beta$, we find the corresponding position of the images $\theta$
numerically in all lens equations (see Section \ref{Numeric} for
details).

A Schwarzschild black hole generates two infinite sequences of
images on each side of the source \cite{Dar}. The outermost pair
of images is made of the classical primary and secondary images,
which can also be described in the classical weak deflection
limit. Section \ref{Sec prisec} is devoted to their study. The
second pair of images is generated by photons performing one
complete loop around the black hole before reaching the observer.
These higher order images are typically very faint, but their
detection could be very important for a confirmation of General
Relativity. We examine them in Section \ref{Sec higord}.

\subsection{Primary and secondary images} \label{Sec prisec}

For any values of $\beta$, we look for solutions of the lens
equation such that $\pi/2<\Phi(\theta,d_{OL},d_{LS})<3\pi/2$. In
this interval we always have one solution for any values of
$\beta$.

\begin{figure}
\resizebox{\hsize}{!}{\includegraphics{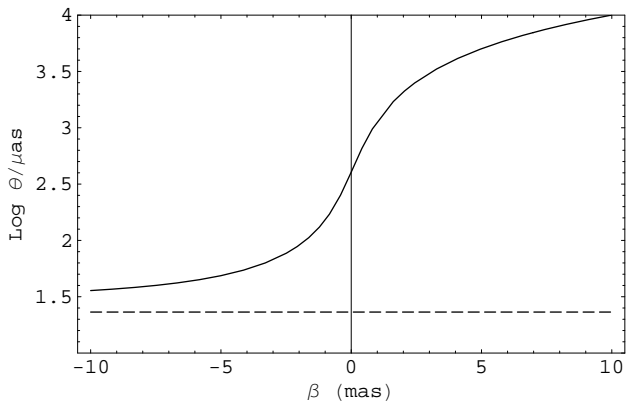}} \caption{The exact
position of the image $\theta$ as a function of the source
position $\beta$ in a linear-log plot. The dashed line is the
shadow of the black hole $\theta_{min}$. For $\beta>0$ the plot
represents the primary image, for $\beta<0$ it represents the
secondary image.}
 \label{Fig theta}
\end{figure}

In Fig. \ref{Fig theta} we show the exact position of the image
for $\beta$ in the range $[-10$ mas$,10$ mas$]$. For positive
$\beta$ the image is on the same side of the source (primary
image), whereas for negative $\beta$ the image is on the opposite
side (secondary image). We see that for large values of $\beta$
the position of the primary image tends to coincide with the
source position itself ($\theta \simeq \beta$). The secondary
image, instead, becomes closer and closer to the black hole. The
minimum angle $\theta_{min}$ represents the border of the
so-called shadow of the black hole. This angle is obtained by the
minimum angular momentum through Eq. (\ref{thetaJ}) with
$J_{min}=3\sqrt{3}r_g/2$. For the black hole in Sgr A*, we have
$\theta_{min}=23$ $\mu$as. The value of $\theta$ for $\beta=0$
represents the radius of the Einstein ring, which for our
geometrical configuration is $\theta_E=404$ $\mu$as.

Fig. \ref{Fig theta} has been obtained with the exact lens
equation. However, as anticipated before, we expect errors at most
of the order $r_g/D_{LS}$ using the approximate equations. So, the
difference cannot be appreciated by superposing the solutions
obtained with different lens equations. We choose, instead, to
plot the relative error in the image positions with respect to the
exact lens equation. These are defined as
\begin{eqnarray}
&& \delta_{OB}=\frac{\theta_{OB}}{\theta_{ex}}-1  \\
&& \delta_{VE}=\frac{\theta_{VE}}{\theta_{ex}}-1  \\
&& \delta_{SA}= \frac{\theta_{SA}}{\theta_{ex}}-1  \\
&& \delta_{WD}= \frac{\theta_{WD}}{\theta_{ex}}-1
\end{eqnarray}
where the subscript ``$ex$'' refers to the result of the exact
lens equation (\ref{LensFKN}), ``$OB$'' refers to the improved
equation presented in this paper (\ref{LensEq}) derived from that
by Ohanian, ``$VE$'' refers to the equation by Virbhadra and Ellis
(\ref{LensVE}), ``$SA$'' refers to the small angles equation
(\ref{LensSA}), and ``$WD$'' refers to the small angles lens
equation with the Einstein approximation $\alpha\simeq 2r_g/D_{OL}
\theta$ for weak deflection angles. Apart from $\theta_{WD}$ and
$\theta_{ex}$, in all remaining cases the position of the images
is computed picking $\alpha$ from Eq. (\ref{alpha}) and relating
$u$ to $\theta$ by Eq. (\ref{thetaJ}), as mentioned before.

\begin{figure}[b]
\resizebox{\hsize}{!}{\includegraphics{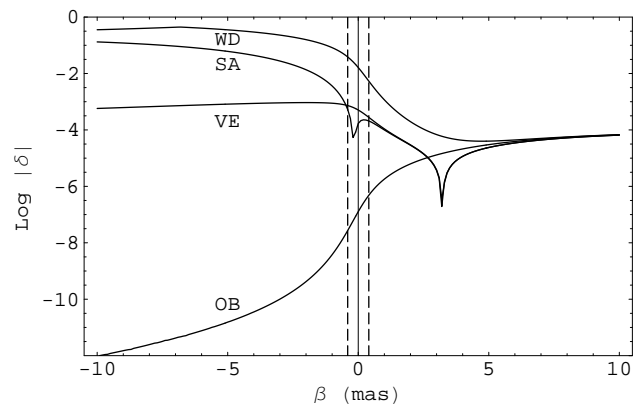}} \caption{Relative
error in the position of the images for the weak deflection lens
equation ($WD$), the small angles lens equation ($SA$), the
Virbhadra and Ellis lens equation ($VE$) and the improved Ohanian
lens equation ($OB$) in a log scale. The dashed vertical lines
bound the region with $|\beta|<\theta_E$, in which gravitational
lensing is mostly interesting.}
 \label{Fig eps}
\end{figure}

Fig. \ref{Fig eps} shows a comparison between the accuracy of the
four equations.

All lens equations have the same accuracy for very large $\beta$,
when gravitational lensing becomes negligible. However, as $\beta$
approaches $\theta_E$ the error of all equations increases, save
for the OB lens equation. Noting that the plot is in a logarithmic
scale, we see that $\delta_{VE}$ tends to be of the order
$10^{-3}$, whereas $\delta_{OB}$ drops to $10^{-7}$, proving to be
much more accurate. Going to $\beta<0$, we see that the error
stays nearly constant for the VE lens equation, whereas it
continues to decrease for the OB lens equation to very tiny
values. Note the change in sign in $\delta_{VE}$ for positive
$\beta$, which can be deduced from the negative spike in the
logarithmic plot as $\delta_{VE}$ crosses zero. This signals the
fact that the sources of error dominating for large $\beta$ and
small $\beta$ are different and have opposite signs. Surprisingly,
the small angles lens equation SA has an error comparable to the
VE lens equation in the most interesting lensing region ($|\beta|
\lesssim \theta_E$), corresponding to a very good alignment of the
source with the lens. The VE lens equation performs better than
the SA equation at large negative $\beta$, where the small angles
approximation is no longer tenable. On the other hand, the weak
deflection approximation WD always stays at larger errors.

Now let us find the interpretation of the errors of the considered
lens equations. We start from the OB lens equation, which only
contains the asymptotic approximation. This lens equation is
completely equivalent to the Ohanian lens equation, so we can use
this equation for the evaluation of the error and translate the
results in terms of the OB equation.

If we expand the azimuthal shift $\Phi$ (\ref{Phi}) in inverse
powers of $D_{OL}$ and $D_{LS}$, at zero order we obtain the
deflection angle $\alpha+\pi$. After this, we obtain some terms
that reproduce the power expansions of the geometrical terms
$-\theta$ and $-\bar\theta$, which are explicitly present in the
Ohanian lens equation (\ref{LensOha}) and finally, the first term
that is not present in the Ohanian lens equation is
\begin{equation}
\Delta \gamma_{asym}=\frac{1}{8}\left[\frac{r_g}{d_{LS}}
\bar\theta^3+ \frac{r_g}{d_{OL}} \theta^3 \right]. \label{asym}
\end{equation}

Taking this term to the right hand side of the Ohanian lens
equation, we can interpret it as an effective change in $\gamma$.
Then we can estimate the corresponding change in $\theta$ simply
by
\begin{equation}
\Delta \theta_{asym}= \frac{\partial \theta}{\partial \gamma}
\Delta \gamma_{asym}.
\end{equation}

\begin{figure}
\resizebox{\hsize}{!}{\includegraphics{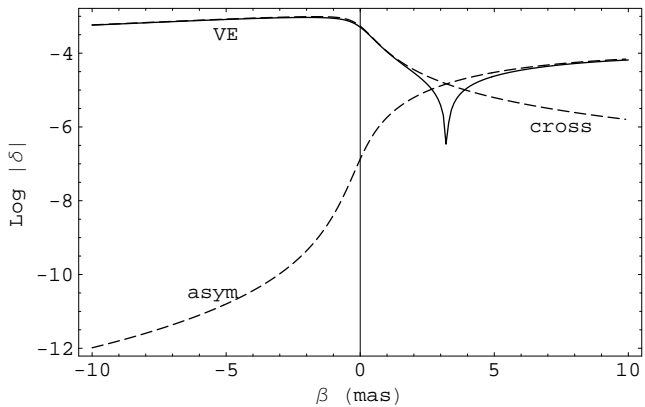}}
\caption{Comparison among the relative error in the position of
the images for the Virbhadra and Ellis lens equation ($VE$) and
what we expect from the asymptotic approximation (asym) and the
approximation that the intersection point $C$ lies on the lens
plane (cross). }
 \label{Fig int}
\end{figure}

Evaluating the derivative of $\theta$ numerically, and taking for
$\Delta \gamma_{asym}$ the expression in Eq. (\ref{asym}), we
exactly reproduce the error $\epsilon_{OB}$ in the OB lens
equation, as can be appreciated from Fig. \ref{Fig int}. In this
figure we have also shown the VE error, already presented in Fig.
\ref{Fig eps}, together with the error deriving from the
approximation that $C$ lies on the lens plane. This error can be
simply calculated by taking the differences of the right hand
sides of Eqs. (\ref{LensVE}) and (\ref{LensEq})
\begin{eqnarray}
&& \Delta \theta_{cross}= \frac{\partial \theta}{\partial \tan
\beta} \Delta (\tan \beta)_{cross} \\
&& \Delta (\tan \beta)_{cross} = \frac{D_{OL}}{D_{OS}}\left[
\tan\theta -\frac{\sin\theta}{\cos(\alpha-\theta)} \right].
\end{eqnarray}

We can explicitly see that the error due to the approximation that
$C$ lies on the lens plane dominates the error due to the
asymptotic approximation in the calculation of the secondary image
($\beta<0$) and for the primary image up to moderately large
$\beta$. For $\beta \gg \theta_E$, the error due to the asymptotic
approximation becomes more relevant than the error due to the
position of the intersection point. However, as already mentioned
before, the last regime is the least important for gravitational
lensing.

The error induced by the small angles approximation can be
obtained by considering the next-to-leading order in the power
expansions of the trigonometric functions present in Eq.
(\ref{LensEq}). It amounts to
\begin{eqnarray}
&&\Delta\beta_{small}=-\frac{\left(D_{OS}\theta-D_{LS}\alpha
\right)^3}{3 D_{OS}^3} \nonumber \\
&& + \frac{D_{OL}\theta (3\alpha^2-6\alpha
\theta+2\theta^2)-2D_{LS}(\alpha-\theta)^3}{6D_{OS}}
\end{eqnarray}

To conclude this subsection, it is interesting to give approximate
estimates of the order of magnitude of the errors in terms of the
geometric distances involved in the problem in different limits.
For example, we can consider the two limits $\beta \gg \theta_E$
(primary image) and $-\beta \gg \theta_E$ (secondary image). In
the first case we have $\theta \simeq \beta$, in the second case
$\theta \simeq -2r_g D_{LS}/(D_{OL}D_{OS} \beta)$ from the weak
deflection approximation. Taking also $D_{OS} \simeq D_{OL}$, we
have the simple estimates
\begin{eqnarray}
&& \delta_{asym+}\simeq -\frac{r_g}{8D_{LS}} \frac{D_{OL}^2
\beta^2}{d_{LS}^2} \\
&& \delta_{asym-}\simeq -\left(\frac{r_g}{D_{OL}\beta} \right)^4
\frac{D_{LS}^2}{d_{LS}^2} \\
&& \delta_{cross+} \simeq \frac{2r_g^2}{D_{OL}^2 \beta^2} \\
&& \delta_{cross-} \simeq \frac{r_g}{D_{LS}} \\
&& \delta_{small+} \simeq -\frac{2r_g^2}{D_{OL}} \\
&& \delta_{small-} \simeq -\frac{D_{OL}^2\beta^2}{3D_{LS}^2},
\end{eqnarray}
which reproduce the curves in Figs. \ref{Fig eps} and \ref{Fig
int} fairly well outside the lensing zone and are useful to
realize the order of magnitude of the errors. In particular, we
can appreciate that the error due to the position of the
intersection point has opposite sign with respect to the others.

Finally, a very interesting limit is the case $\beta=0$,
corresponding to a source perfectly aligned with the black hole
and the two images merged into an Einstein ring of radius
$\theta$. The errors in the estimate of the radius of the Einstein
ring are
\begin{eqnarray}
&& \delta_{asym_0}\simeq -\frac{r_g^2}{8D_{LS}^2} \\
&& \delta_{cross_0} \simeq \frac{r_g}{2D_{LS}} \\
&& \delta_{small_0} \simeq \frac{r_g}{6D_{LS}} \\
&& \delta_{weak_0} \simeq -\frac{15 \pi
}{64}\sqrt{\frac{r_g}{2D_{LS}}},
\end{eqnarray}
where the error due to the weak deflection approximation comes
from the first term neglected in the weak deflection expansion
(see for example Refs. \cite{Sar,KeePet,SerCos}).

These relations are very useful to understand the accuracy of the
various lens equations in the most interesting regime, that is
when the gravitational lensing images are most prominent and
eventually form an Einstein ring. The weak deflection lens
equation has an error of the order $(r_g/D_{LS})^{1/2}$. In the
case under examination, this would translate into an error of 6.6
$\mu$as in the radius of the Einstein ring, which is
$\theta_E=404$ $\mu$as in the situation imagined in this
calculation. The small angles lens equation employed with the
exact deflection angle has an error of order $(r_g/D_{LS})$, which
would amount to 0.07 $\mu$as. The VE lens equation has an error of
the same order of magnitude, but with a slightly larger numerical
coefficient, thus leading to an error of 0.2 $\mu$as. Finally, the
OB lens equation has an error of order $(r_g/D_{LS})^2$, which
amounts to $5\times 10^{-5}$ $\mu$as.

\subsection{Higher order images} \label{Sec higord}

\begin{figure}
\resizebox{\hsize}{!}{\includegraphics{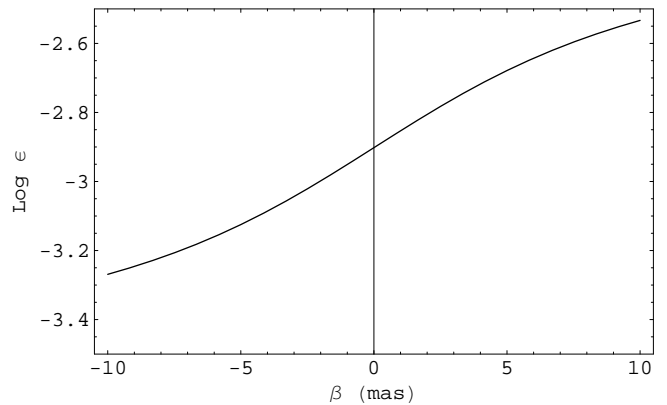}}
\caption{Fractional displacement of the first pair of higher order
images from the shadow border as a function of $\beta$ as defined
in Eq. (\ref{epsilon}). }
 \label{Fig hig}
\end{figure}

An analysis similar to Section \ref{Sec prisec} can be repeated
for the higher order images, whose position and magnification has
been analyzed in Ref. \cite{VirEll} for the case of the black hole
in the center of our Galaxy, and then revisited in many papers
with different methods. The first pair of higher order images can
be found in the interval
$5\pi/2<\Phi(\theta,d_{OL},d_{LS})<7\pi/2$, corresponding to
photons performing one loop before reaching an observer on the
other side of the lens with respect to the source. Since these
images appear very close to the shadow border $\theta_{min}$, we
think it is more instructive to discuss the fractional
displacement of the image from the shadow border, defined as
\begin{equation}
\epsilon=\frac{\theta}{\theta_{min}}-1. \label{epsilon}
\end{equation}
This quantity is shown in Fig. \ref{Fig hig} for the geometric
configuration examined in this paper. The plot has been calculated
using the exact lens equation (\ref{LensFKN}). For $\beta>0$ the
curve represents the fractional displacement of the positive
parity image on the same side of the source and for $\beta<0$ the
curve represents the fractional displacement of the negative
parity image on the opposite side. For $\beta=0$, we obtain the
displacement of the first order Einstein ring $\theta_{E,1}$ from
the shadow border.

\begin{figure}
\resizebox{\hsize}{!}{\includegraphics{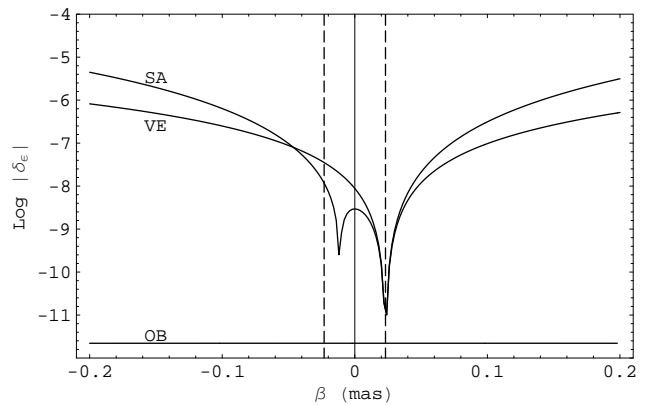}} \caption{Errors in
the estimate of the fractional displacement of the first pair of
higher order images from the shadow borders. $SA$ refers to the
small angles lens equation, $VE$ to the Virbhadra and Ellis lens
equation, $OB$ to the improved Ohanian lens equation. The dashed
lines bound the range in which $|\beta|< \theta_{E,1}$, where
$\theta_{E,1}$ is the angular radius of the first order Einstein
ring.}
 \label{Fig higdelta}
\end{figure}

In order to compare the performances of the different lens
equations, in Fig. \ref{Fig higdelta} we plot the relative error
in this fractional displacement with respect to the exact lens
equation
\begin{eqnarray}
&& \delta_{\epsilon,OB}= \frac{\epsilon_{OB}}{\epsilon_{ex}}-1 \\
&& \delta_{\epsilon,VE}= \frac{\epsilon_{VE}}{\epsilon_{ex}}-1  \\
&& \delta_{\epsilon,SA}= \frac{\epsilon_{SA}}{\epsilon_{ex}}-1.
\end{eqnarray}
The SA lens equation, in this case, is Eq. (\ref{LensSA}) with
$\alpha$ replaced by $\alpha-2\pi$, as explained in Section
\ref{Sec SA}.

Again we see that the OB lens equation has much lower errors in
the estimate of the position of the images than any other lens
equations. The VE lens equation performs slightly better than the
small angles equation for $|\beta|\gtrsim \theta_{E,1}$, whereas
the opposite occurs in the good alignment regime $|\beta|\lesssim
\theta_{E,1}$. It is interesting to note that there is a value for
$\beta$ such that $\alpha=2\pi$ and $\beta=\theta$. In this
particular point, all lens equations are equivalent and are
affected by the same small error, determined by the asymptotic
approximation.

Similarly to the case of the primary and secondary images, we can
make analytical estimates of the errors in the positions of the
higher order images. We follow the same procedure described in the
previous subsection and use the analytical approximation
\begin{equation}
\epsilon\simeq\epsilon_{SDL}=216(7-4\sqrt{3})e^{-3\pi+\gamma},
\end{equation}
derived in the strong deflection limit \cite{Dar,Oha,Boz1}, in
order to have fully analytical results. Then, the errors in the
displacement of the first order Einstein ring are
\begin{eqnarray}
&& \delta_{\epsilon,OB,0}=-\frac{81\sqrt{3}}{64}\left(\frac{r_g}{D_{LS}}\right)^4 \\
&& \delta_{\epsilon,VE,0}=\frac{81\sqrt{3}}{16}\left(\frac{r_g}{D_{LS}}\right)^3 \\
&&
\delta_{\epsilon,SA,0}=\frac{27\sqrt{3}}{16}\left(\frac{r_g}{D_{LS}}\right)^3.
\end{eqnarray}

Again the error in the VE lens equation is of the same order as
the error in the small angles lens equation with a slightly larger
numerical coefficient.

\subsection{Notes on numerical integration} \label{Numeric}

In the previous subsections we have compared the images obtained
by solving different lens equations. The relative error has proved
to be quite small, particularly for the OB lens equation, whose
error goes down to $10^{-12}$ in some plots. In order to
investigate such tiny differences, we must push the numerical
precision of our calculations sufficiently far.

The crucial step in the computations is represented by the
evaluation of the integral (\ref{Phi}), which contains an
integrable singularity at $r_0$. We have used the NIntegrate
routine by Mathematica with the default method (adaptive Gaussian
quadratures with error estimation based on Kronrod points
\cite{Piessens}). In order to reduce the errors, we have worked
with the variable $z$, defined by
\begin{equation}
r=\frac{r_0}{1-z},
\end{equation}
$z$ ranges from 0 to 1 as $r$ ranges from $r_0$ to $+\infty$. In
the Schwarzschild metric, the integrand in Eq. (\ref{Phi})
expressed in terms of $z$ becomes
\begin{equation}
\frac{\sqrt{r_0}}{\sqrt{z}\sqrt{2r_0-3+(3-r_0)z-z^2}}.
\end{equation}

We have checked that increasing the precision goal and the working
precision makes the numerical results converge to an asymptotic
limit, with a controllable precision (we have reached a $10^{-25}$
precision). Furthermore, we have also re-written the integral as
an ordinary differential equation to be solved by the NDSolve
routine, which switches between a non-stiff Adams method and a
stiff Gear backward differentiation formula method. Solving the
differential equation, we get the primitive function, which can be
evaluated at both ends of the domain. In this way, we have
double-checked our results obtained by NIntegrate finding that the
two methods converge to the same result. In particular, we find
that the differential equation method converges more slowly than
the numerical integration with the Gaussian quadratures, as the
precision goal is increased. For this reason we tend to give a
preference to the NIntegrate routine. The results presented in the
plots are calculated with a $10^{-18}$ precision.

The accuracy in the numerical calculation is finally confirmed by
the very good agreement with the analytical estimates for the
errors at leading order.

\section{Discussion and conclusions}

Of course, an approximation can be valid or not depending on the
precision of the observational instrumentation, the level of
environmental noise and the accuracy needed for the extraction of
the interesting information.

The best resolution we can expect to reach in a reasonably near
future is that of the MAXIM project (http://maxim.gsfc.nasa.gov),
which amounts to a fraction of $\mu$as. Then, in the example
considered in this section, the weak deflection lens equation
would not be able to ensure such a precision. The VE and the small
angles lens equations would be marginally adequate, whereas the
Ohanian lens equation would be largely sufficient, without need to
resort to the exact lens equation. So, in order to exploit the
resolution of future instruments, it might be necessary to use the
Ohanian lens equation or one of its variants.

Environmental noise can be provided by scattering and absorption
processes affecting the photons involved in gravitational lensing.
Such processes should be very effective in the environment of the
supermassive black hole at the Galactic center. However, they act
in a statistical way and a good average process should be able to
restore the original information regarding gravitational lensing.
More subtle are the effects that introduce unwanted systematics,
such as gravitational lensing by secondary objects. However, the
gravitational lensing optical depth towards the center of the
bulge is of the order $10^{-6}$ \cite{MolRou}, which indicates
that lensing or microlensing effects by stars or other compact
objects on the line of sight is absolutely negligible (see also
\cite{CGM}). In the absence of relevant systematics, the position
of the centroids of the images can be used for a clean
reconstruction of the gravitational lensing event at the angular
resolution level reached by the instruments.

Finally, we note that there are many theoretical reasons to
require a reconstruction of a gravitational lensing event as
precise as possible. As anticipated, the study of higher order
terms in the deflection angle requires a precision better than
$(r_g/D_{LS})^{1/2}$ at second PPN order and $(r_g/D_{LS})$ at
third PPN order. Therefore, if one wants to compare the results of
PPN formalism with different black hole metrics, it is mandatory
to use a lens equation that guarantees the necessary accuracy. In
this respect the VE lens equation does not represent an
improvement with respect to the small angles approximation, as it
is affected by an error of the same order of magnitude.

In this work we have compared the level of accuracy of several
lens equations, keeping the exact general relativistic lens
equation introduced by Frittelli, Kling and Newman
\cite{FriNew,FKN} as the reference equation. We have shown that
the Ohanian lens equation \cite{Oha} and its close relatives
\cite{BozMan,BozSer} are the best approximations of the exact lens
equation, in that they adopt the asymptotic approximation only.
Other lens equations are not expressed in terms of relative
positions of source, lens and observer \cite{VNC} or introduce
additional approximations \cite{VirEll,DabSch}.

Furthermore, we have presented a new formulation of the Ohanian
lens equation in terms of the distances between the observer, lens
and source planes, which fills a gap in the lens equation zoo. As
shown by a numerical example describing a realistic gravitational
lensing event by the black hole at the center of our Galaxy, such
an equation represents a noteworthy improvement with respect to
previous commonly used lens equations expressed in terms of the
same quantities.

\begin{acknowledgments}
The author thanks Gaetano Scarpetta and Mauro Sereno for useful
comments on the manuscript. We acknowledge support for this work
by MIUR through PRIN 2006 Protocol 2006023491\_003, by research
funds of Agenzia Spaziale Italiana, and by research funds of
Salerno University.
\end{acknowledgments}

\end{document}